\documentclass[aps, prb, reprint, showpacs]{revtex4-1}

\usepackage{graphicx}% Include figure files
\usepackage{dcolumn}% Align table columns on decimal point
\usepackage{bm}% bold math
\usepackage{wrapfig}
\usepackage{scalefnt}

\usepackage{array,booktabs,tabularx,tabulary}
\usepackage{lipsum}
\newcolumntype{L}{>{\raggedright\arraybackslash}X}% <-- added
      \usepackage{siunitx}% <-- added
      \usepackage{caption}% <-- added
\usepackage[export]{adjustbox}
\usepackage[colorlinks=true, linkcolor=red, citecolor=blue, urlcolor=blue, linktoc=page, bookmarks=false, pdfstartview={FitH}, pdfborder={0 0 0.0 [3 3]}]{hyperref}

\usepackage{amsmath}
\usepackage{natbib}
\usepackage{natmove}
\begin{document} 

\title{Electronic and magnetic properties of low dimensional system Co$_2$TeO$_3$Cl$_2$ }

%_________AUTHOR_________
\author{Jayita Chakraborty}
\email[Email: ]{Jayita.Chakraborty1@gmail.com}
\affiliation{Department of Physics, Indian Institute of Science Education and Research Bhopal, Bhauri, Bhopal 462066, India}

\date{\today}% It is always \today, today,

\begin{abstract}
The electronic and magnetic properties of transition metal oxyhalide compound Co$_2$TeO$_3$Cl$_2$ is investigated using first principle calculations within  the framework of density functional theory. In order to find underlying spin lattice of this  compound, various hopping integrals and exchange interactions  are calculated. The calculations reveal that the dominant inter-chain and intra-chain interactions are in $ab$ plane. The exchange path is visualised by Wannier function plotting. The nearest neighbour and next nearest neighbour exchange interactions are antiferromagnetic, making the system frustrated in low dimension.  The importance of spin orbit coupling in this compound is also investigated. The spin quantization axis is favoured along  the crystallographic $b$ direction.  
 \end{abstract}

\pacs{}% PACS, the Physics and Astronomy
\maketitle
\section{Introduction}	
%The electronic and magnetic properties of transition metal oxide materials are largely influenced by dimensionality  and lattice geometry. 

Low dimensional spin systems have attracted much attention over the past few decades due to their variety of unusual magnetic properties.\cite{lemmens2003, BrayPRL1975, JohannesPRB06,  rosnercdcu2b2o6, JayitaPRB2012} Effective low dimensionality means anisotropic exchange interactions in different direction of the crystal that arises due to interplay of geometry and chemical bonding. Due to enhanced quantum fluctuations in low dimension, exotic properties  can be seen in these class of materials. Presence of magnetic frustration in low dimensional spin systems  is particularly important. 
Many complex compounds with different spin lattices like  spin ladder, kagome lattice, triangular lattice, spin chain  show fascinating  magnetic properties like spin gap state, spin liquid state, various type of complex magnetic order etc.\cite{Fak2012, Jeong2011, kotikagome, kotibsco, Chakraborty2016} Transition metal oxyhalide compounds comprising a p-element having a stereochemically active lone pair (such as Te$^{4+}$, Se$^{4+}$, Sb$^{3+}$  etc.), form an attractive field  of study. There is high probability of finding new low-dimensional and frustrated spin system in these class of materials.  	
%The compounds with formula M-Te-O-X (M= Cu, Ni, Fe,Co, X=Cl, Br) are  synthesized recently. These systems show low dimensionality and These systems contain highly polarizable lone pair cation Te$^{4+}$ (5$s^2$5$p^0$). 
The example of such systems  are: FeTe$_2$O$_5$Br,  Cu$_2$Te$_2$O$_5$Cl$_2$, Ni$_5$(TeO$_3$)$_4$Cl$_2$, etc.\cite{beckerjacs2006, Crowe2006, 		Johnsson2003} Due 	to reduced dimensionality and geometric frustration, these materials often show complex magnetic order and interesting magnetic properties. Some of these materials also show multiferroic property.\cite{pregeljprl2009, ChakrabortyPRB13} Number of such oxyhalide compounds are synthesised and studied over time. However, the family of Co$^{2+}$ containing oxohalogenides including a lone-pair cation is not studied much. Co$_2$TeO$_3$Cl$_2$ is one  such compound which is synthesized by Becker {\em{et al}}.\cite{Becker2006} It has a layered structure. The magnetic properties measurements by them suggests long range antiferromagnetic interactions below 30 K and the Curie Weiss temperature is $-$97 K.\cite{Becker2006}  Neutron diffraction study of a powder sample suggests that Co spins are parallel to the crystallographic $b$ directions.\cite{Kashi2007} Exact spin lattice of the system is not known. In complex crystal structures, the microscopic understanding of the spin lattice remains a challenging problem. The  knowledge of intra-chain and inter-chain exchange interactions are  crucial to understand the magnetic behavior of these materials. {\em ab initio} calculations play important role in this direction.
 
 In this paper, the electronic and magnetic properties  of Co$_2$TeO$_3$Cl$_2$ are studied using first principle calculations within the framework of density functional theory (DFT). To determine the underlying spin lattice of this compound, various hopping integrals and exchange interactions between Co$^{2+}$ ions are calculated. In order to assess the importance of  spin orbit coupling (SOC) in this system, the electronic structure calculations are also carried out with SOC and magnetic anisotropy energy is calculated. 
 The remainder of the paper is organized as follows: the crystal structure and computational details are discussed in section II. The results of  electronic structure calculations is discussed in section III. Finally the  results are summarised in section IV.

%Due to the presence of stereochemically active lone-pair and the halide ion contribute to open up the crystal structures by terminating structural elements and by occupying space in the nonbonding regions and increase the possibilities for low dimensional arrangements. This.

\section {\label{sec:crystal} Crystal Structure and Computational details}
%%%%%%%%%%%%%%%%%%%%%%%%%%%%%---Fig.1---%%%%%%%%%%%%%%%%%%%%%%%%%%%%%%%%%%%
\begin{figure}
\centering
\includegraphics[width=0.5\textwidth]{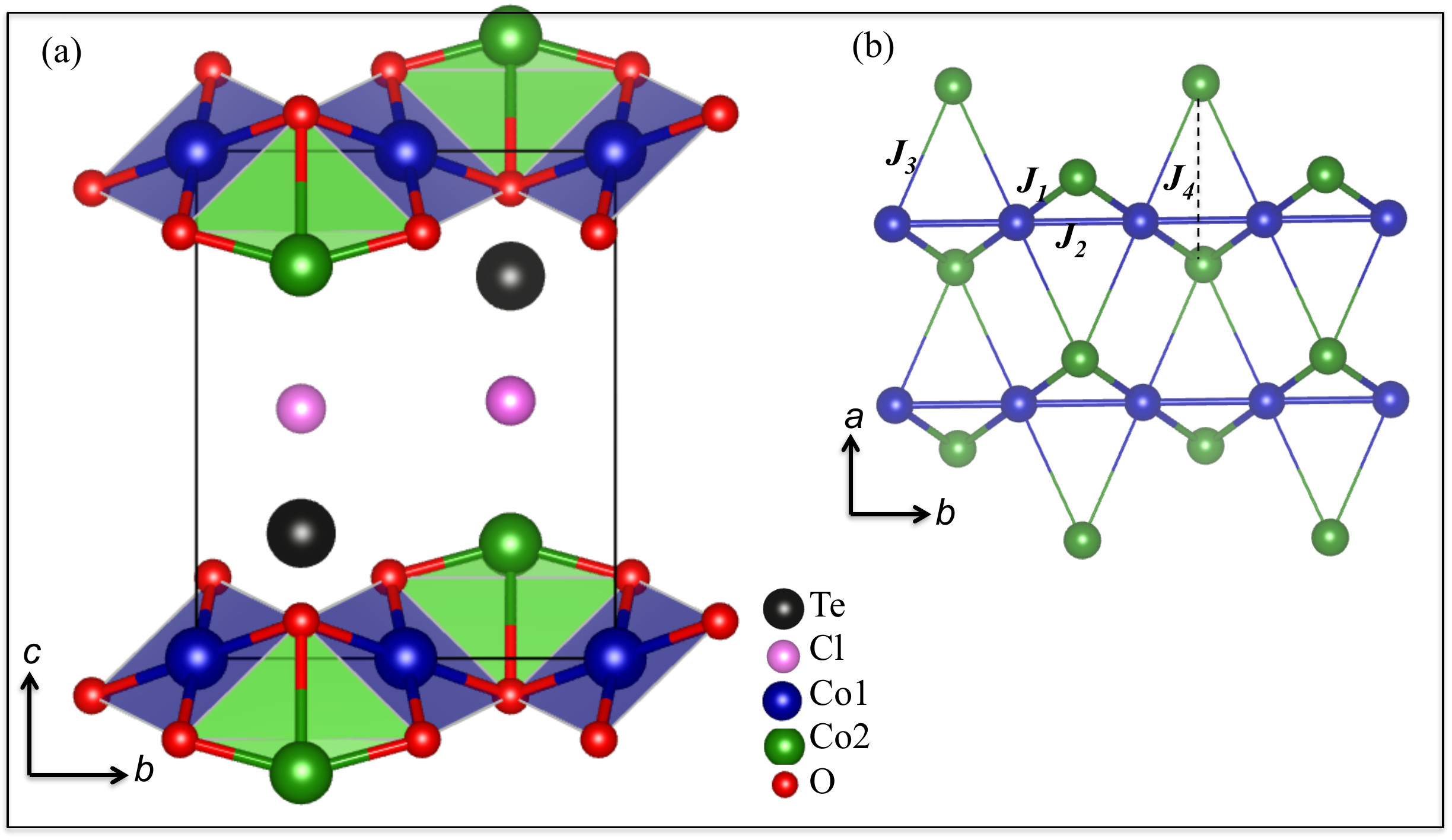}
\caption{\label{fig:structure} (a) The unit cell of Co$_2$TeO$_3$Cl$_2$. (b) The exchange paths}
\end{figure}
%%%%%%%%%%%%%%%%%%%%%%%%%%%%%%%%%%%%%%%%%%%%%%%%%%%%%%%%%%%%%%%%%%%%%%%%%%%
The compound Co$_2$TeO$_3$Cl$_2$ (see Fig.~\ref{fig:structure}) has monoclinic structure with space group P21/m (No. 11). The lattice parameters are $a$ = 5.0472 \AA, $b$=6.6325 \AA, $c$= 8.3452 \AA, and $\beta$=105.43$^\circ$.\cite{Becker2006}  There are two crystallographically different Co ions in this system. Co1 is in square planar environment with two oxygen ions at a distance of 2.053 \AA~ and other two oxygen ions at a distance of 2.084 \AA. Further Co1 is coordinated by two chlorine atoms at a distance 2.497 \AA,   forming Co1O$_4$Cl$_2$  octahedron.
The Co2 atom is coordinated by three oxygen atoms and to three chlorine atoms and forming Co2O$_3$Cl$_3$ octahedron. The Co1O$_4$Cl$_2$  octahedra  share edges with Co2O$_3$Cl$_3$ octahedra and form layered structure. These layers are connected by Te ions. In a layer, the nearest Co1-Co2 distance is 3.06 \AA. Fig.~\ref{fig:structure}(b) shows the arrangement of Co ions in a layer. 
 
%\begin{wrapfigure}{l}{0.5\textwidth}
%\includegraphics[width=1\linewidth]{structure} 
%\caption{Caption1}
%\label{fig:structure}
%\end{wrapfigure}

In the present study, electronic structure calculations are carried out using two different basis sets: (i) the muffin-tin orbital based linearized muffin-tin orbital (LMTO) method  within atomic sphere approximation (ASA) and N$^{\rm th}$ order muffin-tin orbital (NMTO) downfolding method as implemented in the Stuttgart code\cite{Anderson} and (ii)  the plane wave based projector augmented wave (PAW)\cite{blochl} method as implemented in the {\scshape vasp} code \cite{vasp}. For the TB-LMTO-ASA calculation, the space filling in the ASA is obtained by inserting appropriate empty spheres in the interstitial regions. The basis set for the self consistent electronic structure calculation for Co$_2$TeO$_3$Cl$_2$  in the LMTO method includes Co ($s$, $p$, $d$),  Te ($s$, $p$), Cl ($s$, $p$), and  O ($s$, $p$) and the rest are downfolded. The self consistency is achieved by Brillouin-zone integrations in a $(8\times8\times8)$ ${\bf k}$ point mesh. NMTO downfolding method, where potentials are taken from self consistent LMTO method, is used to derive the low energy Hamiltonian. In this method, the low energy model Hamiltonian is constructed by selective downfolding method via integration process. The high energy degrees of freedom are integrated out from the all orbital calculation.  

For the  calculations using plane wave based method, the wave functions are expanded in the plane-wave basis with a kinetic energy cutoff of 500 eV.  The exchange-correlation functional is chosen to be the Perdew-Burke-Ernzerhof (PBE) implementation of the generalized gradient approximation (GGA).\cite{gga} To properly describe the electron correlation associated with the 3$d$ states of Co, GGA+U method is used in the calculations.\cite{dudarev1998}  For electronic structure calculations with spin-orbit coupling (SOC), it is included in scalar relativistic form as a perturbation to the original Hamiltonian. The structural relaxations are carried out until the Hellman-Feynman forces became less than 0.01 eV/\AA. The calculation revealed that there is  very small deviation of the atomic position  in the relaxed structure from experimental structure. Therefore all the calculations reported here have been done with experimental lattice parameters and atomic positions.

%To find out the low energy model, we have employed the $N^{th}$ order muffin-tin orbital (NMTO) downfolding method \cite{Saha-Dasgupta,Tank,Saha}. This method generates the basis set which describes an isolated band or group of bands. The low energy model Hamiltonian is constructed by selective downfolding method via integration process. The high energy degrees of freedom are integrated out from the all orbital calculation.  

%All the calculations reported here have been done with experimental lattice parameters and atomic positions. As the experimental atomic positions are not available for  Ca$_3$ZnCoO$_6$, electronic structure calculations for this compound are carried out with the atomic positions available for Ca$_3$ZnMnO$_6$. This choice is based on the fact that the experimental lattice parameters for Ca$_3$ZnCoO$_6$ are very similar to Ca$_3$ZnMnO$_6$ and the ionic radii of Co$^{4+}$ and Mn$^{4+}$ are nearly identical. This was further corroborated by performing a calculation for  Ca$_3$ZnCoO$_6$ where atomic positions of the ions in the unit cell were allowed to relax keeping the lattice parameters fixed to the experimental values. The calculation revealed very small deviation of the atomic position from that reported for  the Ca$_3$ZnMnO$_6$ compound.

%********************************************************************
\section{\label{sec:results}Results and Discussions}
\subsection {Spin-unpolarized calculation}
%%%%%%%%%%%%%%%%%%%%%%%%%%%%%%%%%%%%%%%%----Fig2----%%%%%%%%%%%%%%%%%%%%%%%%%%%%%%%
\begin{figure}
\centering
\includegraphics[width=0.5\textwidth, right]{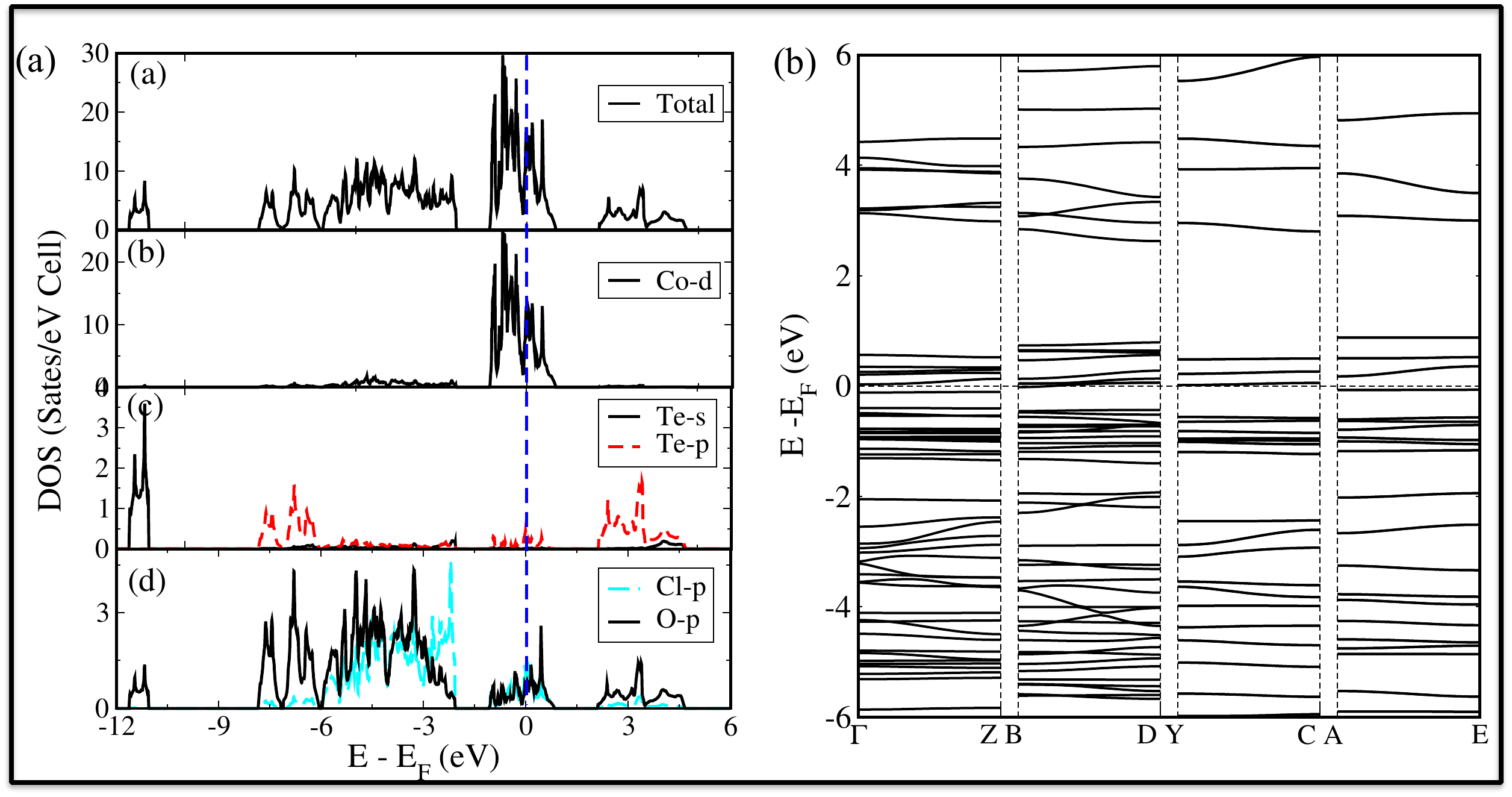}
\caption{\label{fig:banddos} (a) Spin-upolarized total and orbital decomposed density of states for Co$_2$TeO$_3$Cl$_2$ (b) LDA bandstructure}
\end {figure}
%%%%%%%%%%%%%%%%%%%%%%%%%%%%%%%%%%%%%%%%%%%%%%%%%%%%%%%%%%%%%%%%%%%%%%%%%%%%%%%%%%%%%%
To begin with, the electronic structure of Co$_2$TeO$_3$Cl$_2$  is studied without magnetic order. The total and orbital projected density of states (DoS) are displayed in Fig.~\ref{fig:banddos}(a).  Non-magnetic DoS reveals that the Fermi level (E$_F$) is dominated by partially filled Co-$d$ orbitals. The O-$p$ and Cl-$p$ states are fully occupied. The O-$p$ states overlap with Co-$d$ states indicating the strong hybridization of Co-$d$ states with  O-$p$ states. As expected, the occupied Te-$s$ states lie far below the Fermi level and empty Te-$p$ states lie above E$_F$, spreading in the energy range of 2 to 5 eV.   The  Te-5$s$ and Te-5$p$ states are strongly hybridized with the O-$p$ states, which in turn hybridize with Co-$d$ states near the Fermi level.  The system is metallic for the spin unpolarized calculation within  local density approximation (LDA). 

The spin unpolarized band structure calculated within LMTO method  is displayed in Fig.~\ref{fig:banddos}(b). The bands are plotted at the various high symmetry points of the Brillouin zone corresponding to the monoclinic  lattice. The Fermi level is dominated by Co-$d$ states hybridized with O-$p$ and Cl-$p$ states. 
\subsection{Crystal Field Splitting and Magnetism}
 %%%%%%%%%%%%%%%%%%%%%%%%
\begin{figure}
\centering
\includegraphics[width=0.5\textwidth]{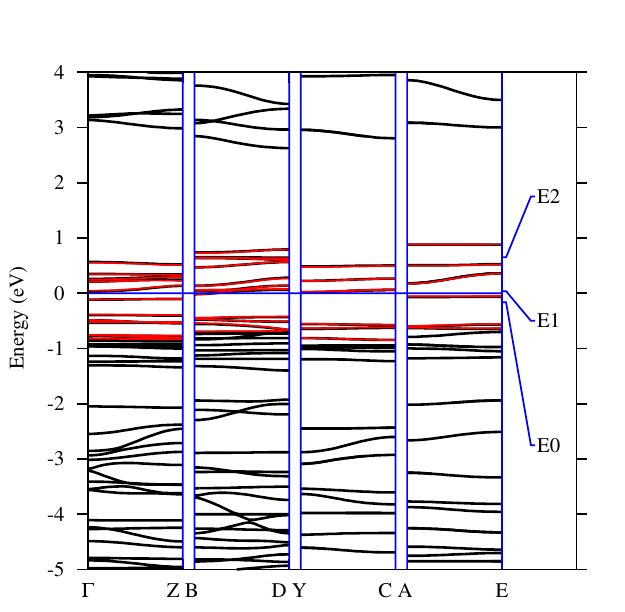}
\caption{\label{fig:nmtofit} The downfolded bandstructure (red line) compared with full LDA bandstructure (black line) of Co$_2$TeO$_3$Cl$_2$}
\end {figure}
%%%%%%%%%%%%%%%%%%%%%%%%
%%%%%%%%%%%%%%%%%%%%%%%%%%%%%%%%%%%
%%%%%%%%%%%%%%%%%%%%%%%%%%%%%%%%%%%%%%%%%%%
 \begin{wrapfigure}{l}{0.3\textwidth}
\includegraphics[width=1\linewidth]{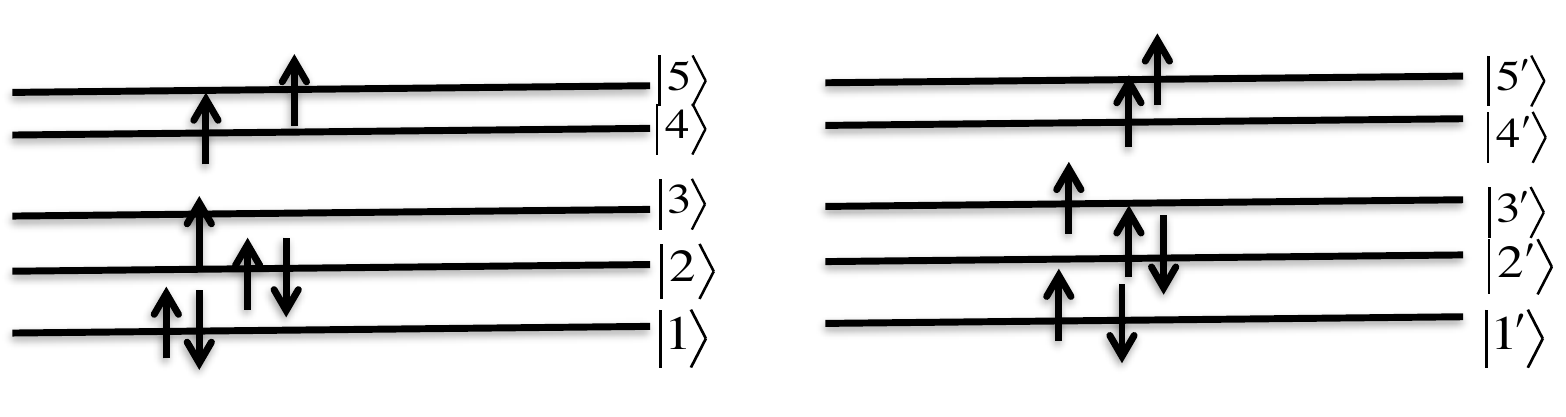} 
\caption{Crystal field splitting and filling of electrons (considering spin polarization) for Co1 and Co2} 
\label{fig:leveldiagram}
\end{wrapfigure}
 
 The NMTO downfolding calculations are performed in order to find the crystal field splitting of Co-$d$ orbitals and hopping interactions between Co ions. To calculate crystal field splitting, only Co-5$d$ orbitals are retained in the basis and the rest  are downfolded. The diagonalization of the onsite block of 5 $\times$ 5 Hamiltonian gives the crystal field splitting for Co1-$d$ and Co2-$d$ states including the covalency with oxygens and chlorines.  These energies are calculated to be $(-2.76, -2.69, -2.58, -1.72$ and $-1.41$ eV$)$ and $(-2.63, -2.61, -2.28, -1.83$ and $-1.43$ eV$)$ for Co1 and Co2 respectively. The corresponding eigenstates turn out to be of mixed character and nondegenerate. The eigen states are written in Appendix A. The level diagram of crystal field splitting and the filling of electrons (including spin polarization) in $d$ level of Co1 and Co2 are shown in Fig.~\ref{fig:leveldiagram}.
There are three partially filled orbitals of $d$-level of Co ions, in  the low energy model   only three of  the $d$ orbitals of Co1 and Co2 are retained in the basis. The downfolded band structure compared with all orbital band structure is displayed in Fig.~\ref{fig:nmtofit} The Fourier transformation of the low energy model Hamiltonian, $H_R=\sum_{ij} t_{ij}\left(c_i^{\dagger}c_j + H.C.\right)$ gives the hopping integrals between various Co atoms. The hopping integrals are displayed in Table~\ref{tab:hop2}. The next nearest neighbour (nnn) hopping interaction ($t_2$) turns out to be strongest. The nearest neighbour (nn)  interaction $t_1$, between Co1-Co2, is compared to the nnn hopping interaction. $t_3$ and $t_4$ are smaller compared to $t_1$ and $t_2$ but non-negligible. Other hoppings in $ab$ plane are negligible. The hopping along crystallographic $c$ direction is also negligible. The strength of the intra-chain and inter-chain hoppings indicate that this system is low dimensional. The antiferromagnetic part of the exchange interaction can be calculated from the hopping integrals using the expression\cite{kugel}:
	\begin{equation}
J^{\rm AF}= \frac{\sum_{ij}{t^2_{ij}}}{U+\Delta_{nn}}	
	\end{equation}
where, $U$ is the Coulomb interaction and $\Delta_{nn}$ is the	onsite energy difference. $U$ = 4 eV and $\Delta_{nn}$ = 0.3 eV are taken in the calculation.   The antiferromagnetic part of the exchange interactions are tabulated in the last column  of Table~\ref{tab:exchange}.
		
%%%%%%%%%%%%%%%%%%%%%%%%%%%%%%%%%%%
\begin{table}
\begin{center}
\caption{\label{tab:hop2} Hopping integrals (in meV) obtained from NMTO downfolding method for Co$_2$TeO$_3$Cl$_2$. The interaction paths (t$_1$, t$_2$ and t$_3$, t$_4$) are indicated in Fig. 1(b) }
%\begin{indented}
%\begin{ruledtabular}
\begin{tabular}{ccccccc}
%\br
\hline
Hopping  & Atoms & Distance  & \multicolumn{4}{c}{orbital involved} \\
 Ing. &  & (\AA) & \multicolumn{4}{c}{} \\
 \hline
 & & &  & $\vert 1 \rangle $ & $\vert 2 \rangle $& $\vert 3\rangle $\\
 $t_1$ & Co1-Co2  & 3.01 & $\langle 1^\prime \vert $ & -22.6 & 0 & -63.3  \\
 
    &     &     &$\langle 2^\prime \vert$ & 28 & 15.4 & 30\\
     &     &     &$\langle 3^\prime \vert$ &-38.4 & -80.2 & -5.1   \\
\hline
 & & &  & $\vert 1 \rangle $ & $\vert 2 \rangle $& $\vert 3\rangle $\\
 $t_2$ & Co1-Co1  & 3.31 & $\langle 1 \vert $ & 47.8 & 23.1 & 23 \\ 
 & & & $\langle 2 \vert$ & 23.1 & -42.8 &  -42.9 \\ 
 & & & $\langle 3 \vert$ & -23 & 42.9 & -83.8\\
\hline
  & & &  & $\vert 1 \rangle $ & $\vert 2 \rangle $& $\vert 3\rangle $\\  
  $t_3$ & Co1-Co2 & 4.12 &  $\langle 1^\prime \vert $ & -19.3 & -29.3  & 21.1\\
    &     &     &$\langle 2^\prime \vert$ & -31.7 & -58.5 & 54.2 \\
     &     &     &$\langle 3^\prime \vert$ & -11.3 & 43.3 & 25\\
\hline
   & & &  & $\vert 1^\prime \rangle $ & $\vert 2^\prime \rangle $& $\vert 3^\prime\rangle $\\         
  $t_4$ & Co2-Co2 & 5.05 &  $\langle 1^\prime \vert $ &  -19.3  & -31.7 &  -11.3 \\   
    &     &     &$\langle 2^\prime \vert$ &  -29.3 &    -58.5 & 43.3 \\ 
    &     &     &$\langle 3^\prime \vert$ &    21.1 &     54.2 &  25\\  
\hline
\end{tabular}
%\end{indented}
%\end{ruledtabular}
\end{center}
\end{table}
%%%%%%%%%%%%%%%%%%%%%%%%%%%%%%%%%%%%%%%%%%%
%%%%%%%%%%%%%%%%%%%%%%%%%%%%%%
\begin{figure}
\centering
\includegraphics[width=0.3\textwidth]{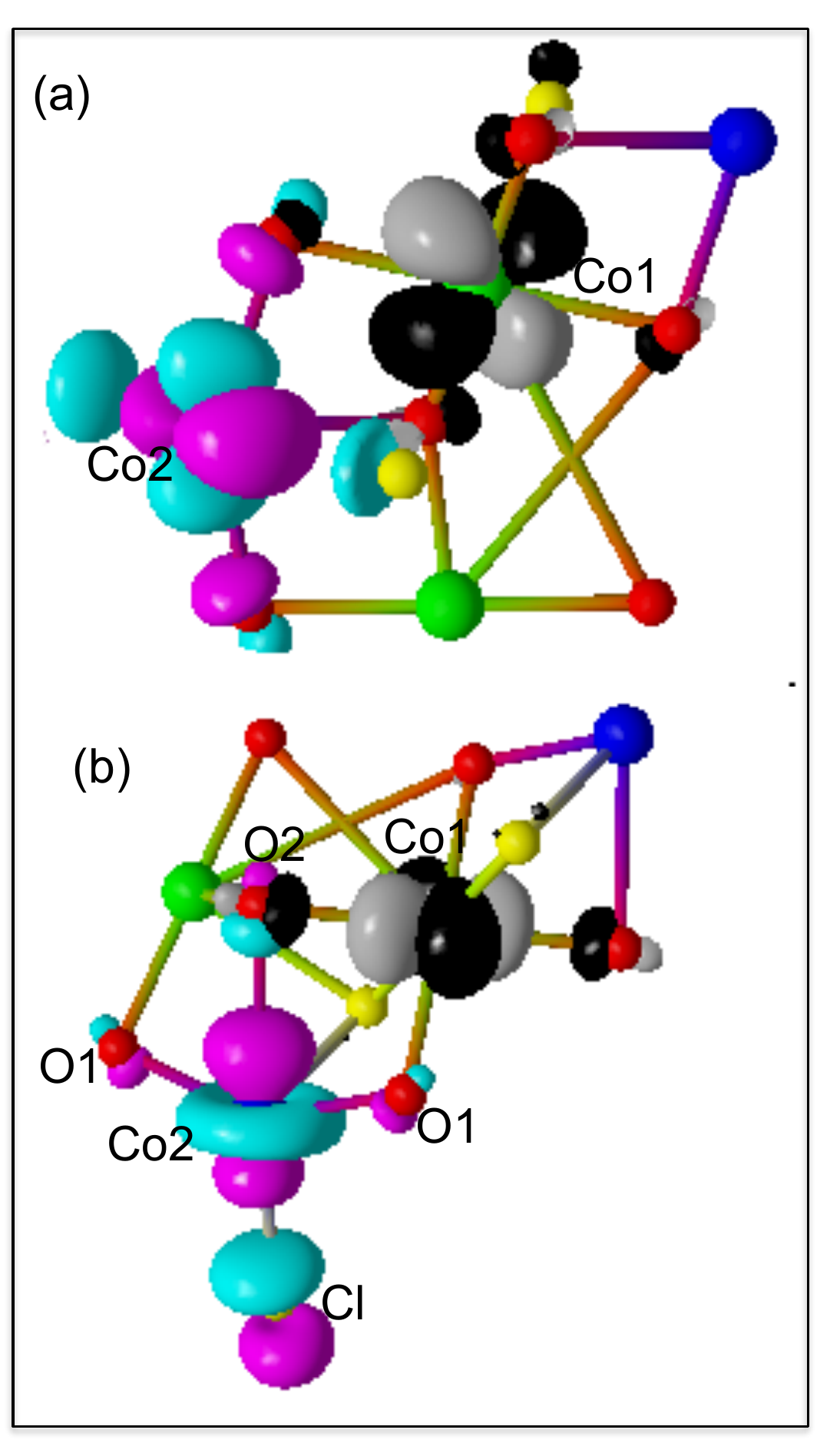}
\caption{\label{fig:wannier} Plot of Wannier functions placed at two different Co sites. The two oppositely signed lobes of the wave functions at site 1 (2), are colored differently as black (magenta) and grey (cyan). (a) Overlap of Co1-$yz$ orbital with Co2-$x^2-y^2$ and (b) Overlap of Co1-$xy$ orbital with Co2-$3z^2-1$ }
\end {figure}
%%%%%%%%%%%%%%%%%%%%%%%%%%%%
In order to visualise the dominant interactions paths, the Wannier functions corresponding to the nearest neighbour Co1 and Co2 ions are plotted in Fig.~\ref{fig:wannier}.   Fig.~\ref{fig:wannier}(a) reveals that the overlap between Co1- $d_{xy}$ effective Wannier functions and that between Co2-$d_{x^2-y^2}$  Wannier functions, placed at two nearest-neighbor Co sites within the chain. These two Co atoms are connected by three oxygens. The super exchange path of $t_1$ hopping is Co1-O-Co2.   Additionally they are connected also by Cl atom. Each Co $d$ oribitals  form strong $pd\sigma$ antibonding linkages with the neighboring O$p_x$/$p_y$ orbitals. The tails of the two Co Wannier functions overlap at the Cl site connecting two Co atoms, forming the crucial exchange path. Wannier function of Co2-$d_{3z^2-r^2}$  is plotted in Fig.~\ref{fig:wannier}(b).

\subsection{Spin polarized Calculations}
\begin{figure}
\centering
\includegraphics[width=0.5\textwidth]{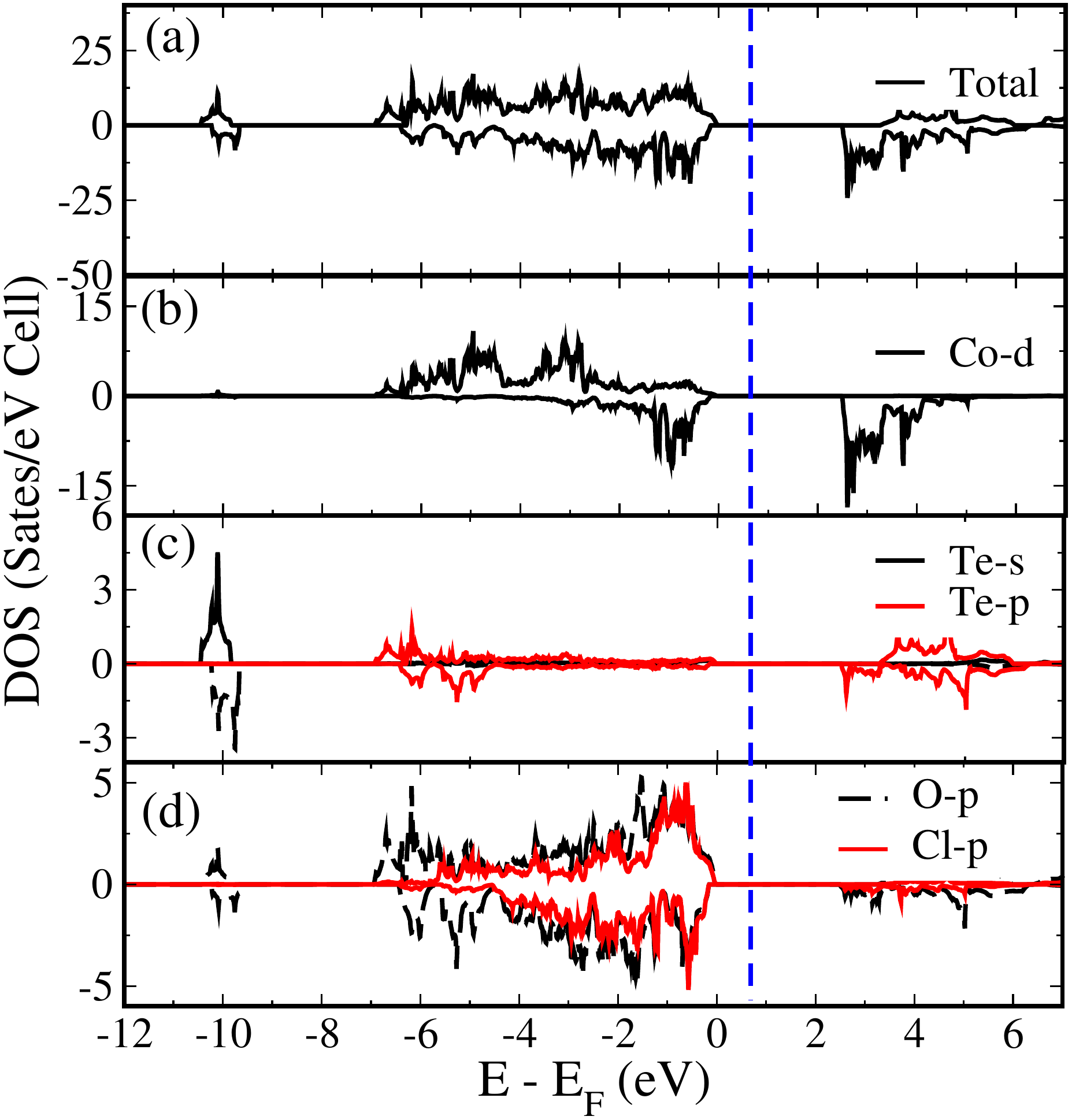}
\caption{\label{fig:dos} Spin polarized total and orbital decomposed density of states for Co$_2$TeO$_3$Cl$_2$}
\end {figure}
Next, the electronic structure of Co$_2$TeO$_3$Cl$_2$ has been  investigated with magnetic order. Fig.~\ref{fig:dos} displayed the spin polarized total and orbital decomposed density of states within GGA+U calculation with $U_{\text{eff}}$= 4 eV. DoS reveals that the majority spin of Co-$d$ states are completely occupied and minority spin states are partially occupied and this picture is consistent with high spin state of Co$^{2+}$ ion. The magnetic moment of Co$^{2+}$ ion is found to be 2.71 $\mu_B$ consistent with high spin configuration of Co$^{2+}$ ion within GGA+U calculation.  The Te-$s$ states are completely occupied and lies far below the Fermi energy. The O-$p$ and Cl-$p$ states are spread over the energy range $-$6 to 6 eV.

\begin{figure}
\centering
\includegraphics[width=0.4\textwidth]{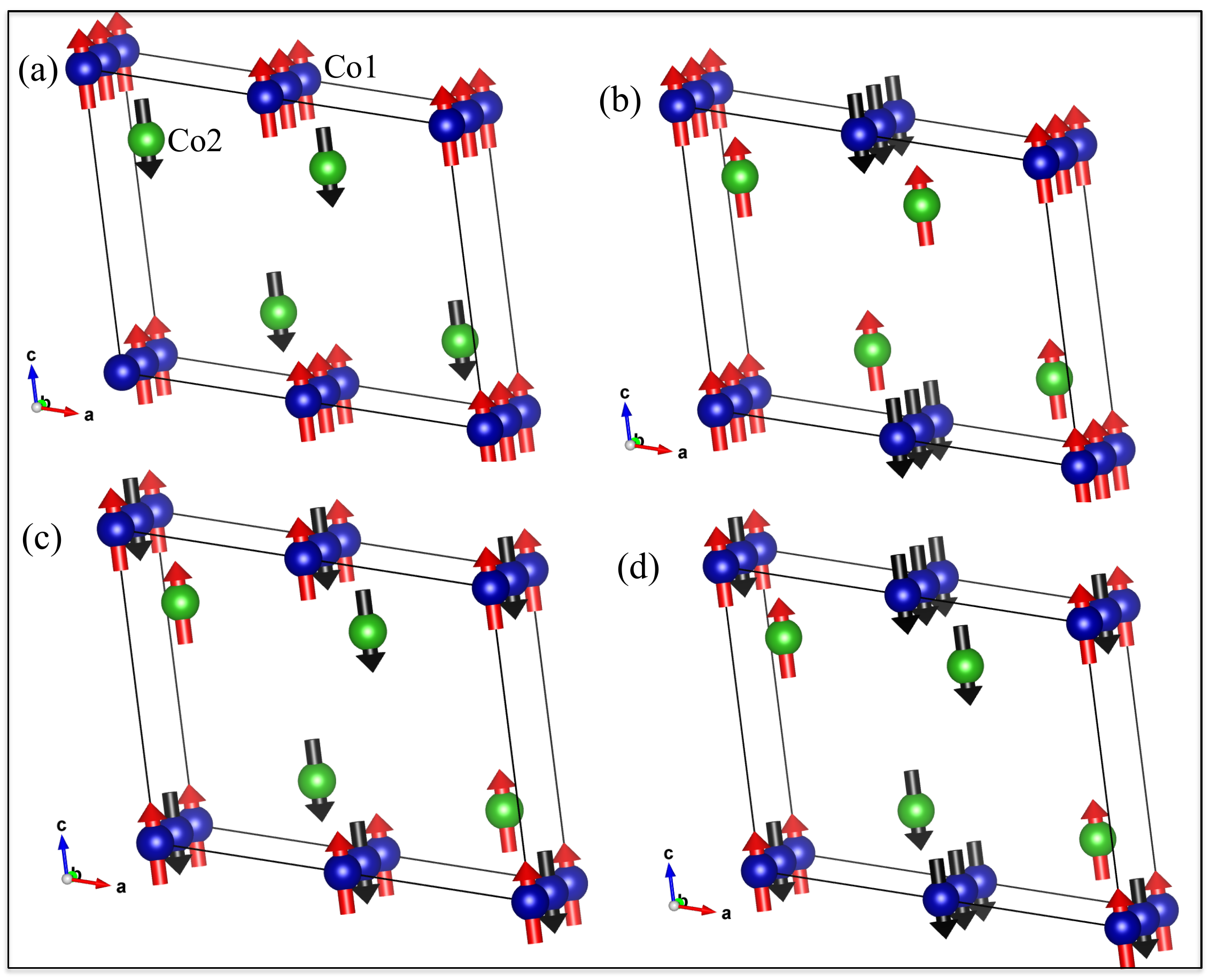}
\caption{\label{fig:magnetic} Different antiferromagnetic configurations (a) AF1, (b) AF2, (c) AF3, and (d) AF4 of Co$_2$TeO$_3$Cl$_2$}
\end {figure}

In the present case, the nearest neighbour exchange interaction is between edge sharing Co1 octahedron with  Co2 octahedron and the angle is  $\angle$Co1-O5-Co2 =81.93$^\circ$. Whereas nnn exchange interaction is between  two corner sharing Co1 octahedra and the angle is $\angle$Co1-O5-Co1 = 105.5$^\circ$.  Since the angles involved in the exchange paths are much deviated from 90$^\circ$ (or 180$^\circ$), Goodenough-Kanamori-Anderson (GKA) rule\cite{book:gka, Kanamori1959} can not be applied here. The nature of interaction between $d^7$-$d^7$ orbital mainly governed by the exchange path and angle with the bridging oxygen atom. In order to find out total magnetic exchange interactions between Co$^{2+}$ ions, total energy calculation method is adopted here.\cite{XiangPRB2007, JayitaPRB2012,Chakraborty2016,kotibsco}  In this method, the total energies  are calculated for several ordered spin configurations of a system and then related the energy differences between these configurations to the corresponding energy differences expected from the Heisenberg spin Hamiltonian:$H = - \sum_{i,j}{\large J_{ij}}{\vec {S}_i}\cdot{\vec {S}_j}$. Here we have used  plane wave basis set. From the hopping strengths, one can conclude that there are  four dominant exchange interactions for this  system as indicated in Fig.~\ref{fig:structure}.
Five different magnetic configurations (see Fig.\ref{fig:magnetic}) are made to find out four dominant exchange interactions. 2$\times$1$\times$1 supercell is used to make different antiferromagnetic arrangements. In AF1 configuration,  Co1 spins are antiferromagnetically coupled with  Co2 spins , whereas in AF2 configuration, Co1 spins are antiparallel along $a$ direction and   all four Co2 spins are parallel with each other.  In AF3 configuration, Co2 spins are antiparallel along $a$ and $c$ directions and Co1 spins are antiparallel only along $b$ direction. AF4 configuration is similar to AF3 configuration except,	  Co1 spins are aligned antiferromagnetically  along $a$ and $b$ directions in AF4 configuration. The energy difference of antiferromagnetic configurations with ferromagnetic configuration, band gap, magnetic moments in each configurations  are displayed in Table~\ref{tab:energy}. The calculated exchange energies are listed in Table~\ref{tab:exchange}.

\begin{table}
\begin{center}
\caption{\label{tab:energy} The relative energies per unit cell, magnetic moments, and band gaps for different magnetic configurations are listed here.}
\begin{ruledtabular}
\begin{tabular}{ccccc}
Magnetic & Band gap & $M_{Co1}$ & $M_{Co2 }$ & Energy\\
Config. & (eV) & ($\mu_B$) & ($\mu_B$) & Difference (meV)\\
\hline
FM & 2.1 & 2.65 & 2.66 & 0\\
AF1 & 2.1  & 2.65 & 2.66 & $-$78.5\\  
AF2 & 2.2 & 2.67 & 2.66  & $-$125\\  
AF3 & 2.2 & 2.67 & 2.63 & $-$33.5\\  
AF4 & 2.2  & 2.67 & 2.6 & $-$135\\
\hline
 \end{tabular}
\end{ruledtabular}
\end{center}
\end{table}

\begin{table}
\begin{center}
\caption{\label{tab:exchange} The exchange interactions (meV) are listed here }
\begin{ruledtabular}
\begin{tabular}{ccccc}
 Exchange & Distance  & Type of Co  & Total & $J^{\rm AF}$ \\
 path &  (\AA)  &  atom & Exchange &  from NMTO\\
\hline
$J_1$  & 3.01 & Co1-Co2  & $-$6.01 & $-$7.3\\
& & intra-chain & & \\
$J_2$ & 3.31 & Co1-Co1 & $-$7.12 & $-$8.7\\
& & intra-chain & & \\
$J_3$  & 4.12 & Co1-Co2  & 3.83 & $-$2.4\\
& & inter-chain & & \\
$J_4$ &  5.05 & Co2-Co2 &$-$4.76 & $-$3.7\\
& & inter-chain & & \\
\end{tabular}
\end{ruledtabular}
\end{center}
\end{table}
The result of the calculations reveal that the nearest neighbour (nn) interaction $J_1$ between Co1 and Co2 is antiferromagnetic. Next nearest neighbour (nnn) interaction between Co1-Co2 is also antiferromagnetic and stronger than $J_1$ interaction. $J_3$ is the inter chain exchange interaction between  Co1-Co2 and it is ferromagnetic type. The inter chain exchange interaction $J_4$ between Co2-Co2 is also antiferromagnetic and comparable to intra chain exchange interactions. Such strong antiferromagnetic interactions make the system frustrated in low dimension. 

\subsection{Spin orbit coupling}
\begin{table}
\begin{center}
\caption{\label{tab:SOC} The spin (orbital) moment of Co$^{2+}$ ions, energy difference (in meV)  within GGA+U+SOC calculations (with $U_{eff}=4$ eV) ion axis are listed here.  }
\begin{ruledtabular}
\begin{tabular}{cccc}
Spin Quantization Axis & Co1 moment ($\mu_B$) & 
Co2 moment & $\Delta E$  \\
$(001)$ &  2.7 (0.21) & 2.7 (0.22) & 0 \\
$(010)$ & 2.66(0.22) & 2.67(0.22) & -2.20 \\
$(100)$ & 2.66(0.22) & 2.65(0.22) & -1.20 \\
\end	{tabular}
\end{ruledtabular}
\end{center}
\end{table}
 In order to investigate the importance of spin orbit coupling and magnetocrystalline anisotropy, the electronic structure calculations are carried out within GGA+U+SOC as implemented in {\scshape vasp}. The spin-quantization axis was chosen to be parallel to the direction of the crystallographic $c$ axis  as well as perpendicular to the $c$ axis. The spin and orbital moment of Co$^{2+}$ ions  listed in Table~\ref{tab:SOC}.  The value of orbital moment suggests the spin orbit coupling is important in this system.  The  magnetocrystalline anisotropy (listed in Table~\ref{tab:SOC}) is quite strong. The spin quantization is found to be favoured along crystallographic $b$ direction.

\section{Summary and Conclusion}
{\em Ab initio} density functional theory calculations are carried out to study the electronic structure and magnetic properties of oxyhallide compound Co$_2$TeO$_3$Cl$_2$. In order to determine the underlying spin lattice,  various hopping integrals and exchange interactions between Co$^{2+}$ ons are calculated. Hopping integrals calculated using N$^{\rm th}$ order muffin-tin orbital (NMTO) downfolding method reveal that both intra-chain and inter-chain hoppings in $ab$ plane are comparable and therefore the system is low dimensional. The dominant exchange interactions are found to be antiferromagnetic. The nearest neighbour interaction $J_1$ and next nearest neighbour interaction $J_2$ are antiferromagnetic, making the system frustrated. $J_1$ is stronger than $J_1$. Results of the present work suggests that the system is two dimensional with competing magnetic interactions.  The importance of spin orbit coupling in this compound is also investigated. The results of calculations reveal that magnetocrystalline anisotropy is strong in Co$_2$TeO$_3$Cl$_2$ and $(010)$ is the easy axis  as observed experimentally. 

\section{Acknowdegement} I like to thank Dr.\ N.\ Ganguli for fruitful scientific discussions. Use of high performance computing facility of IISER Bhopal is gratefully acknowledged.

%%%%%%%%%%%%%%%%%%%%%%%%%%%%%%%%%%%%%%%%%%%
\section{APPENDIX A}
  The eigen states of Co1 and Co2 obtained from NMTO method are following:
\begin{eqnarray*}
\vert1\rangle  &=&  {\scalefont{2} 0.09\vert {xy} \rangle +0.16} \vert {yz}\rangle + 0.20 \vert {3z^2-1}\rangle + 0.49 \vert {xz} \rangle +0.82 \vert {x^2-y^2} \rangle \\ \nonumber
\vert2\rangle  &=& 0.50\vert {xy}\rangle -  0.49 \vert yz \rangle  + 0.69 \vert 3z^2-1\rangle +0.03 \vert xz \rangle -0.14 \vert x^2-y^2 \rangle   \\ \nonumber
\vert3\rangle  &=&  0.15 \vert xy \rangle -  0.63 \vert yz \rangle - 0.45 \vert 3z^2-r^2\rangle - 0.38 \vert xz \rangle + 0.46 \vert x^2-y^2 \rangle  \\ \nonumber
\vert4\rangle  &=& 0.60 \vert xy \rangle   + 0.56 \vert yz \rangle + 0.02 \vert 3z^2-r^2\rangle - 0.54 \vert xz \rangle + 0.14 \vert x^2-y^2 \rangle  \\ \nonumber
\vert5\rangle &=& 0.59\vert xy \rangle -0.01 \vert yz \rangle  -0.52 \vert 3z^2-r^2\rangle +0.55 \vert xz \rangle -0.26 \vert x^2-y^2 \rangle  \\ \nonumber
\\ \nonumber
 \vert1^\prime\rangle  &=&  -0.93 \vert xy \rangle +0.34 \vert yz\rangle \\ \nonumber
\vert2^\prime\rangle  &=&    0.30 \vert 3z^2-r^2\rangle - 0.92 \vert xz \rangle -0.34 \vert x^2-y^2 \rangle   \\ \nonumber
\vert3^\prime\rangle  &=&  -0.34 \vert xy \rangle -  0.93 \vert yz \rangle \\ \nonumber
\vert4^\prime\rangle  &=&  - 0.38 \vert 3z^2-r^2\rangle - 0.68 \vert xz \rangle + 0.61 \vert x^2-y^2 \rangle  \\ \nonumber
\vert5^\prime\rangle &=&   0.66 \vert 3z^2-r^2\rangle +0.02 \vert xz \rangle + 0.74 \vert x^2-y^2 \rangle  \\ \nonumber  
\end{eqnarray*}


\begin{thebibliography}{26}%
\makeatletter
\providecommand \@ifxundefined [1]{%
 \@ifx{#1\undefined}
}%
\providecommand \@ifnum [1]{%
 \ifnum #1\expandafter \@firstoftwo
 \else \expandafter \@secondoftwo
 \fi
}%
\providecommand \@ifx [1]{%
 \ifx #1\expandafter \@firstoftwo
 \else \expandafter \@secondoftwo
 \fi
}%
\providecommand \natexlab [1]{#1}%
\providecommand \enquote  [1]{``#1''}%
\providecommand \bibnamefont  [1]{#1}%
\providecommand \bibfnamefont [1]{#1}%
\providecommand \citenamefont [1]{#1}%
\providecommand \href@noop [0]{\@secondoftwo}%
\providecommand \href [0]{\begingroup \@sanitize@url \@href}%
\providecommand \@href[1]{\@@startlink{#1}\@@href}%
\providecommand \@@href[1]{\endgroup#1\@@endlink}%
\providecommand \@sanitize@url [0]{\catcode `\\12\catcode `\$12\catcode
  `\&12\catcode `\#12\catcode `\^12\catcode `\_12\catcode `\%12\relax}%
\providecommand \@@startlink[1]{}%
\providecommand \@@endlink[0]{}%
\providecommand \url  [0]{\begingroup\@sanitize@url \@url }%
\providecommand \@url [1]{\endgroup\@href {#1}{\urlprefix }}%
\providecommand \urlprefix  [0]{URL }%
\providecommand \Eprint [0]{\href }%
\providecommand \doibase [0]{http://dx.doi.org/}%
\providecommand \selectlanguage [0]{\@gobble}%
\providecommand \bibinfo  [0]{\@secondoftwo}%
\providecommand \bibfield  [0]{\@secondoftwo}%
\providecommand \translation [1]{[#1]}%
\providecommand \BibitemOpen [0]{}%
\providecommand \bibitemStop [0]{}%
\providecommand \bibitemNoStop [0]{.\EOS\space}%
\providecommand \EOS [0]{\spacefactor3000\relax}%
\providecommand \BibitemShut  [1]{\csname bibitem#1\endcsname}%
\let\auto@bib@innerbib\@empty
%</preamble>
\bibitem [{\citenamefont {Lemmens}\ \emph {et~al.}(2003)\citenamefont
  {Lemmens}, \citenamefont {G{\"{n}}therodt},\ and\ \citenamefont
  {Gros}}]{lemmens2003}%
  \BibitemOpen
  \bibfield  {author} {\bibinfo {author} {\bibfnamefont {P.}~\bibnamefont
  {Lemmens}}, \bibinfo {author} {\bibfnamefont {G.}~\bibnamefont
  {G{\"{n}}therodt}}, \ and\ \bibinfo {author} {\bibfnamefont {C.}~\bibnamefont
  {Gros}},\ }\href {\doibase 10.1016/S0370-1573(02)00321-6} {\bibfield
  {journal} {\bibinfo  {journal} {Phys. Rep.}\ }\textbf {\bibinfo {volume}
  {375}},\ \bibinfo {pages} {1 } (\bibinfo {year} {2003})}\BibitemShut
  {NoStop}%
\bibitem [{\citenamefont {Bray}\ \emph {et~al.}(1975)\citenamefont {Bray},
  \citenamefont {Hart}, \citenamefont {Interrante}, \citenamefont {Jacobs},
  \citenamefont {Kasper}, \citenamefont {Watkins}, \citenamefont {Wee},\ and\
  \citenamefont {Bonner}}]{BrayPRL1975}%
  \BibitemOpen
  \bibfield  {author} {\bibinfo {author} {\bibfnamefont {J.~W.}\ \bibnamefont
  {Bray}}, \bibinfo {author} {\bibfnamefont {H.~R.}\ \bibnamefont {Hart}},
  \bibinfo {author} {\bibfnamefont {L.~V.}\ \bibnamefont {Interrante}},
  \bibinfo {author} {\bibfnamefont {I.~S.}\ \bibnamefont {Jacobs}}, \bibinfo
  {author} {\bibfnamefont {J.~S.}\ \bibnamefont {Kasper}}, \bibinfo {author}
  {\bibfnamefont {G.~D.}\ \bibnamefont {Watkins}}, \bibinfo {author}
  {\bibfnamefont {S.~H.}\ \bibnamefont {Wee}}, \ and\ \bibinfo {author}
  {\bibfnamefont {J.~C.}\ \bibnamefont {Bonner}},\ }\href {\doibase
  10.1103/PhysRevLett.35.744} {\bibfield  {journal} {\bibinfo  {journal} {Phys.
  Rev. Lett.}\ }\textbf {\bibinfo {volume} {35}},\ \bibinfo {pages} {744}
  (\bibinfo {year} {1975})}\BibitemShut {NoStop}%
\bibitem [{\citenamefont {Johannes}\ \emph {et~al.}(2006)\citenamefont
  {Johannes}, \citenamefont {Richter}, \citenamefont {Drechsler},\ and\
  \citenamefont {Rosner}}]{JohannesPRB06}%
  \BibitemOpen
  \bibfield  {author} {\bibinfo {author} {\bibfnamefont {M.~D.}\ \bibnamefont
  {Johannes}}, \bibinfo {author} {\bibfnamefont {J.}~\bibnamefont {Richter}},
  \bibinfo {author} {\bibfnamefont {S.-L.}\ \bibnamefont {Drechsler}}, \ and\
  \bibinfo {author} {\bibfnamefont {H.}~\bibnamefont {Rosner}},\ }\href
  {\doibase 10.1103/PhysRevB.74.174435} {\bibfield  {journal} {\bibinfo
  {journal} {Phys. Rev. B}\ }\textbf {\bibinfo {volume} {74}},\ \bibinfo
  {pages} {174435} (\bibinfo {year} {2006})}\BibitemShut {NoStop}%
\bibitem [{\citenamefont {Janson}\ \emph {et~al.}(2012)\citenamefont {Janson},
  \citenamefont {Rousochatzakis}, \citenamefont {Tsirlin}, \citenamefont
  {Richter}, \citenamefont {Skourski},\ and\ \citenamefont
  {Rosner}}]{rosnercdcu2b2o6}%
  \BibitemOpen
  \bibfield  {author} {\bibinfo {author} {\bibfnamefont {O.}~\bibnamefont
  {Janson}}, \bibinfo {author} {\bibfnamefont {I.}~\bibnamefont
  {Rousochatzakis}}, \bibinfo {author} {\bibfnamefont {A.~A.}\ \bibnamefont
  {Tsirlin}}, \bibinfo {author} {\bibfnamefont {J.}~\bibnamefont {Richter}},
  \bibinfo {author} {\bibfnamefont {Y.}~\bibnamefont {Skourski}}, \ and\
  \bibinfo {author} {\bibfnamefont {H.}~\bibnamefont {Rosner}},\ }\href
  {\doibase 10.1103/PhysRevB.85.064404} {\bibfield  {journal} {\bibinfo
  {journal} {Phys. Rev. B}\ }\textbf {\bibinfo {volume} {85}},\ \bibinfo
  {pages} {064404} (\bibinfo {year} {2012})}\BibitemShut {NoStop}%
\bibitem [{\citenamefont {Chakraborty}\ and\ \citenamefont
  {Dasgupta}(2012)}]{JayitaPRB2012}%
  \BibitemOpen
  \bibfield  {author} {\bibinfo {author} {\bibfnamefont {J.}~\bibnamefont
  {Chakraborty}}\ and\ \bibinfo {author} {\bibfnamefont {I.}~\bibnamefont
  {Dasgupta}},\ }\href {\doibase 10.1103/PhysRevB.86.054434} {\bibfield
  {journal} {\bibinfo  {journal} {Phys. Rev. B}\ }\textbf {\bibinfo {volume}
  {86}},\ \bibinfo {pages} {054434} (\bibinfo {year} {2012})}\BibitemShut
  {NoStop}%
\bibitem [{\citenamefont {F\aa{}k}\ \emph {et~al.}(2012)\citenamefont
  {F\aa{}k}, \citenamefont {Kermarrec}, \citenamefont {Messio}, \citenamefont
  {Bernu}, \citenamefont {Lhuillier}, \citenamefont {Bert}, \citenamefont
  {Mendels}, \citenamefont {Koteswararao}, \citenamefont {Bouquet},
  \citenamefont {Ollivier}, \citenamefont {Hillier}, \citenamefont {Amato},
  \citenamefont {Colman},\ and\ \citenamefont {Wills}}]{Fak2012}%
  \BibitemOpen
  \bibfield  {author} {\bibinfo {author} {\bibfnamefont {B.}~\bibnamefont
  {F\aa{}k}}, \bibinfo {author} {\bibfnamefont {E.}~\bibnamefont {Kermarrec}},
  \bibinfo {author} {\bibfnamefont {L.}~\bibnamefont {Messio}}, \bibinfo
  {author} {\bibfnamefont {B.}~\bibnamefont {Bernu}}, \bibinfo {author}
  {\bibfnamefont {C.}~\bibnamefont {Lhuillier}}, \bibinfo {author}
  {\bibfnamefont {F.}~\bibnamefont {Bert}}, \bibinfo {author} {\bibfnamefont
  {P.}~\bibnamefont {Mendels}}, \bibinfo {author} {\bibfnamefont
  {B.}~\bibnamefont {Koteswararao}}, \bibinfo {author} {\bibfnamefont
  {F.}~\bibnamefont {Bouquet}}, \bibinfo {author} {\bibfnamefont
  {J.}~\bibnamefont {Ollivier}}, \bibinfo {author} {\bibfnamefont {A.~D.}\
  \bibnamefont {Hillier}}, \bibinfo {author} {\bibfnamefont {A.}~\bibnamefont
  {Amato}}, \bibinfo {author} {\bibfnamefont {R.~H.}\ \bibnamefont {Colman}}, \
  and\ \bibinfo {author} {\bibfnamefont {A.~S.}\ \bibnamefont {Wills}},\ }\href
  {\doibase 10.1103/PhysRevLett.109.037208} {\bibfield  {journal} {\bibinfo
  {journal} {Phys. Rev. Lett.}\ }\textbf {\bibinfo {volume} {109}},\ \bibinfo
  {pages} {037208} (\bibinfo {year} {2012})}\BibitemShut {NoStop}%
\bibitem [{\citenamefont {Jeong}\ \emph {et~al.}(2011)\citenamefont {Jeong},
  \citenamefont {Bert}, \citenamefont {Mendels}, \citenamefont {Duc},
  \citenamefont {Trombe}, \citenamefont {De-Vries},\ and\ \citenamefont
  {Harrison}}]{Jeong2011}%
  \BibitemOpen
  \bibfield  {author} {\bibinfo {author} {\bibfnamefont {M.}~\bibnamefont
  {Jeong}}, \bibinfo {author} {\bibfnamefont {F.}~\bibnamefont {Bert}},
  \bibinfo {author} {\bibfnamefont {P.}~\bibnamefont {Mendels}}, \bibinfo
  {author} {\bibfnamefont {F.}~\bibnamefont {Duc}}, \bibinfo {author}
  {\bibfnamefont {J.~C.}\ \bibnamefont {Trombe}}, \bibinfo {author}
  {\bibfnamefont {M.~A.}\ \bibnamefont {De-Vries}}, \ and\ \bibinfo {author}
  {\bibfnamefont {A.}~\bibnamefont {Harrison}},\ }\href {\doibase
  10.1103/PhysRevLett.107.237201} {\bibfield  {journal} {\bibinfo  {journal}
  {Phys. Rev. Lett.}\ }\textbf {\bibinfo {volume} {107}},\ \bibinfo {pages}
  {237201} (\bibinfo {year} {2011})}\BibitemShut {NoStop}%
\bibitem [{\citenamefont {Koteswararao}\ \emph {et~al.}(2013)\citenamefont
  {Koteswararao}, \citenamefont {Kumar}, \citenamefont {Chakraborty},
  \citenamefont {Jeon}, \citenamefont {Mahajan}, \citenamefont {Dasgupta},
  \citenamefont {Kim},\ and\ \citenamefont {Chou}}]{kotikagome}%
  \BibitemOpen
  \bibfield  {author} {\bibinfo {author} {\bibfnamefont {B.}~\bibnamefont
  {Koteswararao}}, \bibinfo {author} {\bibfnamefont {R.}~\bibnamefont {Kumar}},
  \bibinfo {author} {\bibfnamefont {J.}~\bibnamefont {Chakraborty}}, \bibinfo
  {author} {\bibfnamefont {B.-G.}\ \bibnamefont {Jeon}}, \bibinfo {author}
  {\bibfnamefont {A.~V.}\ \bibnamefont {Mahajan}}, \bibinfo {author}
  {\bibfnamefont {I.}~\bibnamefont {Dasgupta}}, \bibinfo {author}
  {\bibfnamefont {K.~H.}\ \bibnamefont {Kim}}, \ and\ \bibinfo {author}
  {\bibfnamefont {F.~C.}\ \bibnamefont {Chou}},\ }\href {\doibase
  10.1088/0953-8984/25/33/336003} {\bibfield  {journal} {\bibinfo  {journal}
  {J. Phys.: Condens. matter}\ }\textbf {\bibinfo {volume} {25}},\ \bibinfo
  {pages} {336003} (\bibinfo {year} {2013})}\BibitemShut {NoStop}%
\bibitem [{\citenamefont {Koteswararao}\ \emph {et~al.}(2012)\citenamefont
  {Koteswararao}, \citenamefont {Mahajan}, \citenamefont {Bert}, \citenamefont
  {Mendels}, \citenamefont {Chakraborty}, \citenamefont {Singh}, \citenamefont
  {Dasgupta}, \citenamefont {Rayaprol}, \citenamefont {Siruguri}, \citenamefont
  {Hoser},\ and\ \citenamefont {Kaushik}}]{kotibsco}%
  \BibitemOpen
  \bibfield  {author} {\bibinfo {author} {\bibfnamefont {B.}~\bibnamefont
  {Koteswararao}}, \bibinfo {author} {\bibfnamefont {A.~V.}\ \bibnamefont
  {Mahajan}}, \bibinfo {author} {\bibfnamefont {F.}~\bibnamefont {Bert}},
  \bibinfo {author} {\bibfnamefont {P.}~\bibnamefont {Mendels}}, \bibinfo
  {author} {\bibfnamefont {J.}~\bibnamefont {Chakraborty}}, \bibinfo {author}
  {\bibfnamefont {V.}~\bibnamefont {Singh}}, \bibinfo {author} {\bibfnamefont
  {I.}~\bibnamefont {Dasgupta}}, \bibinfo {author} {\bibfnamefont
  {S.}~\bibnamefont {Rayaprol}}, \bibinfo {author} {\bibfnamefont
  {V.}~\bibnamefont {Siruguri}}, \bibinfo {author} {\bibfnamefont
  {a.}~\bibnamefont {Hoser}}, \ and\ \bibinfo {author} {\bibfnamefont {S.~D.}\
  \bibnamefont {Kaushik}},\ }\href {\doibase 10.1088/0953-8984/24/23/236001}
  {\bibfield  {journal} {\bibinfo  {journal} {J. Phys.: Condens. Matter}\
  }\textbf {\bibinfo {volume} {24}},\ \bibinfo {pages} {236001} (\bibinfo
  {year} {2012})}\BibitemShut {NoStop}%
\bibitem [{\citenamefont {Chakraborty}\ \emph {et~al.}(2016)\citenamefont
  {Chakraborty}, \citenamefont {Samanta}, \citenamefont {Nanda},\ and\
  \citenamefont {Dasgupta}}]{Chakraborty2016}%
  \BibitemOpen
  \bibfield  {author} {\bibinfo {author} {\bibfnamefont {J.}~\bibnamefont
  {Chakraborty}}, \bibinfo {author} {\bibfnamefont {S.}~\bibnamefont
  {Samanta}}, \bibinfo {author} {\bibfnamefont {B.~R.~K.}\ \bibnamefont
  {Nanda}}, \ and\ \bibinfo {author} {\bibfnamefont {I.}~\bibnamefont
  {Dasgupta}},\ }\href {\doibase 10.1088/0953-8984/28/37/375501} {\bibfield
  {journal} {\bibinfo  {journal} {J. Phys.: Condens. Matter}\ }\textbf
  {\bibinfo {volume} {28}},\ \bibinfo {pages} {375501} (\bibinfo {year}
  {2016})}\BibitemShut {NoStop}%
\bibitem [{\citenamefont {Becker}\ \emph
  {et~al.}(2006{\natexlab{a}})\citenamefont {Becker}, \citenamefont {Johnsson},
  \citenamefont {Kremer}, \citenamefont {Klauss},\ and\ \citenamefont
  {Lemmens}}]{beckerjacs2006}%
  \BibitemOpen
  \bibfield  {author} {\bibinfo {author} {\bibfnamefont {R.}~\bibnamefont
  {Becker}}, \bibinfo {author} {\bibfnamefont {M.}~\bibnamefont {Johnsson}},
  \bibinfo {author} {\bibfnamefont {R.~K.}\ \bibnamefont {Kremer}}, \bibinfo
  {author} {\bibfnamefont {H.-H.}\ \bibnamefont {Klauss}}, \ and\ \bibinfo
  {author} {\bibfnamefont {P.}~\bibnamefont {Lemmens}},\ }\href {\doibase
  10.1021/ja064738d} {\bibfield  {journal} {\bibinfo  {journal} {J. Am. Chem.
  Soc.}\ }\textbf {\bibinfo {volume} {128}},\ \bibinfo {pages} {15469}
  (\bibinfo {year} {2006}{\natexlab{a}})}\BibitemShut {NoStop}%
\bibitem [{\citenamefont {{Rie Takagi, Mats Johnsson, Vladimir Gnezdilov,
  Reinhard K. Kremer, Wolfram Brenig}}\ and\ \citenamefont
  {Lemmens}(2006)}]{Crowe2006}%
  \BibitemOpen
  \bibfield  {author} {\bibinfo {author} {\bibnamefont {{Rie Takagi, Mats
  Johnsson, Vladimir Gnezdilov, Reinhard K. Kremer, Wolfram Brenig}}}\ and\
  \bibinfo {author} {\bibfnamefont {P.}~\bibnamefont {Lemmens}},\ }\href
  {\doibase 10.1103/PhysRevB.74.144413} {\bibfield  {journal} {\bibinfo
  {journal} {Phys. Rev. B}\ }\textbf {\bibinfo {volume} {74}},\ \bibinfo
  {pages} {144413} (\bibinfo {year} {2006})}\BibitemShut {NoStop}%
\bibitem [{\citenamefont {Johnsson}\ \emph {et~al.}(2003)\citenamefont
  {Johnsson}, \citenamefont {T{\"{o}}rnroos}, \citenamefont {Lemmens},\ and\
  \citenamefont {Millet}}]{Johnsson2003}%
  \BibitemOpen
  \bibfield  {author} {\bibinfo {author} {\bibfnamefont {M.}~\bibnamefont
  {Johnsson}}, \bibinfo {author} {\bibfnamefont {K.~W.}\ \bibnamefont
  {T{\"{o}}rnroos}}, \bibinfo {author} {\bibfnamefont {P.}~\bibnamefont
  {Lemmens}}, \ and\ \bibinfo {author} {\bibfnamefont {P.}~\bibnamefont
  {Millet}},\ }\href {\doibase 10.1021/cm0206587} {\bibfield  {journal}
  {\bibinfo  {journal} {Chemistry of Materials}\ }\textbf {\bibinfo {volume}
  {15}},\ \bibinfo {pages} {68} (\bibinfo {year} {2003})}\BibitemShut {NoStop}%
\bibitem [{\citenamefont {Pregelj}\ \emph {et~al.}(2009)\citenamefont
  {Pregelj}, \citenamefont {Zaharko}, \citenamefont {Zorko}, \citenamefont
  {Kutnjak}, \citenamefont {Jegli\ifmmode~\check{c}\else \v{c}\fi{}},
  \citenamefont {Brown}, \citenamefont {Jagodi\ifmmode~\check{c}\else
  \v{c}\fi{}}, \citenamefont {Jagli\ifmmode \check{c}\else
  \v{c}\fi{}i\ifmmode~\acute{c}\else \'{c}\fi{}}, \citenamefont {Berger},\ and\
  \citenamefont {Ar\ifmmode~\check{c}\else \v{c}\fi{}on}}]{pregeljprl2009}%
  \BibitemOpen
  \bibfield  {author} {\bibinfo {author} {\bibfnamefont {M.}~\bibnamefont
  {Pregelj}}, \bibinfo {author} {\bibfnamefont {O.}~\bibnamefont {Zaharko}},
  \bibinfo {author} {\bibfnamefont {A.}~\bibnamefont {Zorko}}, \bibinfo
  {author} {\bibfnamefont {Z.}~\bibnamefont {Kutnjak}}, \bibinfo {author}
  {\bibfnamefont {P.}~\bibnamefont {Jegli\ifmmode~\check{c}\else \v{c}\fi{}}},
  \bibinfo {author} {\bibfnamefont {P.~J.}\ \bibnamefont {Brown}}, \bibinfo
  {author} {\bibfnamefont {M.}~\bibnamefont {Jagodi\ifmmode~\check{c}\else
  \v{c}\fi{}}}, \bibinfo {author} {\bibfnamefont {Z.}~\bibnamefont
  {Jagli\ifmmode \check{c}\else \v{c}\fi{}i\ifmmode~\acute{c}\else
  \'{c}\fi{}}}, \bibinfo {author} {\bibfnamefont {H.}~\bibnamefont {Berger}}, \
  and\ \bibinfo {author} {\bibfnamefont {D.}~\bibnamefont
  {Ar\ifmmode~\check{c}\else \v{c}\fi{}on}},\ }\href {\doibase
  10.1103/PhysRevLett.103.147202} {\bibfield  {journal} {\bibinfo  {journal}
  {Phys. Rev. Lett.}\ }\textbf {\bibinfo {volume} {103}},\ \bibinfo {pages}
  {147202} (\bibinfo {year} {2009})}\BibitemShut {NoStop}%
\bibitem [{\citenamefont {Chakraborty}\ \emph {et~al.}(2013)\citenamefont
  {Chakraborty}, \citenamefont {Ganguli}, \citenamefont {Saha-dasgupta},\ and\
  \citenamefont {Dasgupta}}]{ChakrabortyPRB13}%
  \BibitemOpen
  \bibfield  {author} {\bibinfo {author} {\bibfnamefont {J.}~\bibnamefont
  {Chakraborty}}, \bibinfo {author} {\bibfnamefont {N.}~\bibnamefont
  {Ganguli}}, \bibinfo {author} {\bibfnamefont {T.}~\bibnamefont
  {Saha-dasgupta}}, \ and\ \bibinfo {author} {\bibfnamefont {I.}~\bibnamefont
  {Dasgupta}},\ }\href {\doibase 10.1103/PhysRevB.88.094409} {\bibfield
  {journal} {\bibinfo  {journal} {Phys. Rev. B}\ }\textbf {\bibinfo {volume}
  {88}},\ \bibinfo {pages} {094409} (\bibinfo {year} {2013})}\BibitemShut
  {NoStop}%
\bibitem [{\citenamefont {Becker}\ \emph
  {et~al.}(2006{\natexlab{b}})\citenamefont {Becker}, \citenamefont {Berger},
  \citenamefont {Johnsson}, \citenamefont {Prester}, \citenamefont {Marohnic},
  \citenamefont {Miljak},\ and\ \citenamefont {Herak}}]{Becker2006}%
  \BibitemOpen
  \bibfield  {author} {\bibinfo {author} {\bibfnamefont {R.}~\bibnamefont
  {Becker}}, \bibinfo {author} {\bibfnamefont {H.}~\bibnamefont {Berger}},
  \bibinfo {author} {\bibfnamefont {M.}~\bibnamefont {Johnsson}}, \bibinfo
  {author} {\bibfnamefont {M.}~\bibnamefont {Prester}}, \bibinfo {author}
  {\bibfnamefont {Z.}~\bibnamefont {Marohnic}}, \bibinfo {author}
  {\bibfnamefont {M.}~\bibnamefont {Miljak}}, \ and\ \bibinfo {author}
  {\bibfnamefont {M.}~\bibnamefont {Herak}},\ }\href {\doibase
  10.1016/j.jssc.2005.12.007} {\bibfield  {journal} {\bibinfo  {journal} {J.
  Solid State Chem.}\ }\textbf {\bibinfo {volume} {179}},\ \bibinfo {pages}
  {836} (\bibinfo {year} {2006}{\natexlab{b}})}\BibitemShut {NoStop}%
\bibitem [{\citenamefont {Kashi}\ \emph {et~al.}(2007)\citenamefont {Kashi},
  \citenamefont {Yasui}, \citenamefont {Moyoshi}, \citenamefont {Sato},
  \citenamefont {Igawa},\ and\ \citenamefont {Kakurai}}]{Kashi2007}%
  \BibitemOpen
  \bibfield  {author} {\bibinfo {author} {\bibfnamefont {T.}~\bibnamefont
  {Kashi}}, \bibinfo {author} {\bibfnamefont {Y.}~\bibnamefont {Yasui}},
  \bibinfo {author} {\bibfnamefont {T.}~\bibnamefont {Moyoshi}}, \bibinfo
  {author} {\bibfnamefont {M.}~\bibnamefont {Sato}}, \bibinfo {author}
  {\bibfnamefont {N.}~\bibnamefont {Igawa}}, \ and\ \bibinfo {author}
  {\bibfnamefont {K.}~\bibnamefont {Kakurai}},\ }\href {\doibase
  10.1143/JPSJ.76.084713} {\bibfield  {journal} {\bibinfo  {journal} {J. Phys.
  Soc. of Japan}\ }\textbf {\bibinfo {volume} {76}},\ \bibinfo {pages} {084713}
  (\bibinfo {year} {2007})}\BibitemShut {NoStop}%
\bibitem [{\citenamefont {Andersen}\ and\ \citenamefont
  {Jepsen}(1984)}]{Anderson}%
  \BibitemOpen
  \bibfield  {author} {\bibinfo {author} {\bibfnamefont {O.~K.}\ \bibnamefont
  {Andersen}}\ and\ \bibinfo {author} {\bibfnamefont {O.}~\bibnamefont
  {Jepsen}},\ }\href {\doibase 10.1103/PhysRevLett.53.2571} {\bibfield
  {journal} {\bibinfo  {journal} {Phys. Rev. Lett.}\ }\textbf {\bibinfo
  {volume} {53}},\ \bibinfo {pages} {2571} (\bibinfo {year}
  {1984})}\BibitemShut {NoStop}%
\bibitem [{\citenamefont {Bl\"ochl}(1994)}]{blochl}%
  \BibitemOpen
  \bibfield  {author} {\bibinfo {author} {\bibfnamefont {P.~E.}\ \bibnamefont
  {Bl\"ochl}},\ }\href {\doibase 10.1103/PhysRevB.50.17953} {\bibfield
  {journal} {\bibinfo  {journal} {Phys. Rev. B}\ }\textbf {\bibinfo {volume}
  {50}},\ \bibinfo {pages} {17953} (\bibinfo {year} {1994})}\BibitemShut
  {NoStop}%
\bibitem [{\citenamefont {Kresse}\ and\ \citenamefont
  {Furthm\"uller}(1996)}]{vasp}%
  \BibitemOpen
  \bibfield  {author} {\bibinfo {author} {\bibfnamefont {G.}~\bibnamefont
  {Kresse}}\ and\ \bibinfo {author} {\bibfnamefont {J.}~\bibnamefont
  {Furthm\"uller}},\ }\href {\doibase 10.1103/PhysRevB.54.11169} {\bibfield
  {journal} {\bibinfo  {journal} {Phys. Rev. B}\ }\textbf {\bibinfo {volume}
  {54}},\ \bibinfo {pages} {11169} (\bibinfo {year} {1996})}\BibitemShut
  {NoStop}%
\bibitem [{\citenamefont {Perdew}\ \emph {et~al.}(1996)\citenamefont {Perdew},
  \citenamefont {Burke},\ and\ \citenamefont {Ernzerhof}}]{gga}%
  \BibitemOpen
  \bibfield  {author} {\bibinfo {author} {\bibfnamefont {J.~P.}\ \bibnamefont
  {Perdew}}, \bibinfo {author} {\bibfnamefont {K.}~\bibnamefont {Burke}}, \
  and\ \bibinfo {author} {\bibfnamefont {M.}~\bibnamefont {Ernzerhof}},\ }\href
  {\doibase 10.1103/PhysRevLett.77.3865} {\bibfield  {journal} {\bibinfo
  {journal} {Phys. Rev. Lett.}\ }\textbf {\bibinfo {volume} {77}},\ \bibinfo
  {pages} {3865} (\bibinfo {year} {1996})}\BibitemShut {NoStop}%
\bibitem [{\citenamefont {Dudarev}\ \emph {et~al.}(1998)\citenamefont
  {Dudarev}, \citenamefont {Botton}, \citenamefont {Savrasov}, \citenamefont
  {Humphreys},\ and\ \citenamefont {Sutton}}]{dudarev1998}%
  \BibitemOpen
  \bibfield  {author} {\bibinfo {author} {\bibfnamefont {S.~L.}\ \bibnamefont
  {Dudarev}}, \bibinfo {author} {\bibfnamefont {G.~A.}\ \bibnamefont {Botton}},
  \bibinfo {author} {\bibfnamefont {S.~Y.}\ \bibnamefont {Savrasov}}, \bibinfo
  {author} {\bibfnamefont {C.~J.}\ \bibnamefont {Humphreys}}, \ and\ \bibinfo
  {author} {\bibfnamefont {A.~P.}\ \bibnamefont {Sutton}},\ }\href {\doibase
  10.1103/PhysRevB.57.1505} {\bibfield  {journal} {\bibinfo  {journal} {Phys.
  Rev. B}\ }\textbf {\bibinfo {volume} {57}},\ \bibinfo {pages} {1505}
  (\bibinfo {year} {1998})}\BibitemShut {NoStop}%
\bibitem [{\citenamefont {Kugel}\ and\ \citenamefont {Khomskii}(1982)}]{kugel}%
  \BibitemOpen
  \bibfield  {author} {\bibinfo {author} {\bibfnamefont {K.~I.}\ \bibnamefont
  {Kugel}}\ and\ \bibinfo {author} {\bibfnamefont {D.~I.}\ \bibnamefont
  {Khomskii}},\ }\href {\doibase 10.1070/PU1982v025n04ABEH004537} {\bibfield
  {journal} {\bibinfo  {journal} {Sov. Phys. Usp.}\ }\textbf {\bibinfo {volume}
  {25}},\ \bibinfo {pages} {231} (\bibinfo {year} {1982})}\BibitemShut
  {NoStop}%
\bibitem [{\citenamefont {Goodenough}(1963)}]{book:gka}%
  \BibitemOpen
  \bibfield  {author} {\bibinfo {author} {\bibfnamefont {J.}~\bibnamefont
  {Goodenough}},\ }\href@noop {} {\emph {\bibinfo {title} {Magnetism and the
  Chemical Bond}}}\ (\bibinfo  {publisher} {New York: Interscience},\ \bibinfo
  {year} {1963})\BibitemShut {NoStop}%
\bibitem [{\citenamefont {Kanamori}(1959)}]{Kanamori1959}%
  \BibitemOpen
  \bibfield  {author} {\bibinfo {author} {\bibfnamefont {J.}~\bibnamefont
  {Kanamori}},\ }\href {\doibase 10.1016/0022-3697(59)90061-7} {\bibfield
  {journal} {\bibinfo  {journal} {J. Phys. Chem. Solids}\ }\textbf {\bibinfo
  {volume} {10}},\ \bibinfo {pages} {87 } (\bibinfo {year} {1959})}\BibitemShut
  {NoStop}%
\bibitem [{\citenamefont {Xiang}\ \emph {et~al.}(2007)\citenamefont {Xiang},
  \citenamefont {Lee},\ and\ \citenamefont {Whangbo}}]{XiangPRB2007}%
  \BibitemOpen
  \bibfield  {author} {\bibinfo {author} {\bibfnamefont {H.~J.}\ \bibnamefont
  {Xiang}}, \bibinfo {author} {\bibfnamefont {C.}~\bibnamefont {Lee}}, \ and\
  \bibinfo {author} {\bibfnamefont {M.~H.}\ \bibnamefont {Whangbo}},\ }\href
  {\doibase 10.1103/PhysRevB.76.220411} {\bibfield  {journal} {\bibinfo
  {journal} {Phys. Rev. B}\ }\textbf {\bibinfo {volume} {76}},\ \bibinfo
  {pages} {220411(R)} (\bibinfo {year} {2007})}\BibitemShut {NoStop}%

\end{thebibliography}
\end{document}